\documentclass[a4paper,11pt]{article}
\usepackage{jcappub}
\usepackage[T1]{fontenc}
\usepackage{subfig}
\usepackage{hyperref}
\usepackage{amsmath}
\usepackage{mathtools}
\usepackage{tablefootnote}
\usepackage{soul}
\usepackage{ulem}
\usepackage{color}
\usepackage{float}
\usepackage{makecell}
\usepackage[usenames,dvipsnames,svgnames,table]{xcolor}
\usepackage{indentfirst}
\usepackage{leftidx}
\usepackage[compat=1.1.0]{tikz-feynman}
\usepackage{multirow}
\definecolor{LightCyan}{rgb}{0.88,1,1}
\newcolumntype{a}{>{\columncolor{LightCyan}}c}

\begin{document}

\title{Distinguishing Dirac vs. Majorana neutrinos: a cosmological probe}

\author[1]{Beatriz Hernandez-Molinero,}
\author[2,3]{Raul Jimenez,}
\author[1,4]{Carlos Pe\~na Garay}

\affiliation[1]{Laboratorio Subterr\'aneo de Canfranc, 22880 - Estaci\'on de Canfranc, Huesca, Spain.}
\affiliation[2]{Institute of Cosmos Sciences, University of Barcelona, Marti  i Franques 1, E-08028 Barcelona, Spain.}
\affiliation[3]{Instituci\`o Catalana de Recerca i Estudis Avan\c{c}ats, Pg. Lluis Companys 23, Barcelona, E-08010, Spain.}
\affiliation[4]{I2SysBio, CSIC-University of Valencia, 46071 - Valencia, Spain.}

\emailAdd{bhernandez@lsc-canfranc.es;raul.jimenez@icc.ub.edu;cpenya@lsc-canfranc.es}

\abstract{
Cosmic background neutrinos ($C_{\nu}B)$ helicity composition is different for Dirac or Majorana neutrinos making detectors based on $C_{\nu}B$ capture sensitive to the nature of neutrinos. We calculate, for the first time, the helicity changes of neutrinos crossing dark matter fields, to quantitatively calculate this effect on the capture rate. We show that a fraction of neutrinos change their helicity, regardless of them being deflected by a void or a dark matter halo. The average signal from the 100 most massive voids or halos in a Gpc$^3$ gives a prediction that if neutrinos are Dirac, the density of the $C_{\nu} B$ background measured on Earth should be 48 cm${^{-3}}$ for left-helical neutrinos, a decrease of  15\% (53.6 cm${^{-3}}$; 5\%) for a halo (void) with respect to the standard calculation without including gravitational effects due to large scale structures. In terms of the total capture rate in a 100 g tritium detector, this translates in $4.9^{+1.1}_{-0.8}$ neutrinos per year for the Dirac case, as a function of the unknown neutrino mass scale, or 8.1 per year if neutrinos are Majorana. Thus although smaller than the factor two for the non-relativistic case, it is still large enough to be detected and it highlights the power of future $C_{\nu} B$ detectors, as an alternative to neutrinoless double beta decay experiments, to discover the neutrino nature.}

\maketitle	
\section{Introduction}

It is now established, experimentally, that neutrinos undergo flavor oscillations~\cite{ParticleDataGroup:2020ssz}. This is so if they are massive particles and there is a mixing matrix between the flavor and mass eigenstates. Because they are neutral and massive fermions, neutrinos could be either Dirac or Majorana even in the presence of only the standard model couplings. If neutrinos are Majorana, then they  violate lepton number and have only two degrees of freedom, distinguished by helicity, for each flavor. Dirac neutrinos have four degrees of freedom for each flavor as all other fermions.

One of the most pressing open questions in physics is to discover if the neutrino is its own anti-particle. The most natural path to discover this is via the presence of neutrinoless double beta decay (which is only possible if the neutrinos are Majorana)\cite{Goeppert-Mayer,furry}; several experimental efforts around the world, placed in underground laboratories, are trying to detect this signal \cite{Giuliani:2019uno}. However, if the total mass of neutrinos is as small as cosmology suggests ($< 0.1$ eV~\cite{Planck:2018vyg,shapefit}), then the required size of the detector will be $\sim$1-10 tons of the double beta decay isotope, which will be challenging. It is therefore interesting to explore other experimental signatures to detect the physical nature of neutrinos. 

A promising avenue is the possibility to explore signatures in the sky for processes which count differently Dirac and Majorana neutrinos and therefore would show that the measurement of the cosmic neutrino background ($C_{\nu} B$) is sensitive to the nature of neutrinos \cite{Long_2014}. There is currently an important ongoing experimental effort to detect the $C_{\nu} B$~\cite{Ptolemy} which will eventually measure and characterize this background. Using this signal and the clustering of dark matter in the sky, we calculate an observational signature that can lead to an astrophysical track of lepton number violation. In particular, under gravity, the deflection angle of neutrinos falling into a gravitational potential modifies the helicity content of Dirac neutrinos in amounts that can not match the helicity content of Majorana neutrinos.

To compute the expected signal-to-noise in the sky we use high-resolution numerical simulations of large scale structure formation that include cold dark matter and massive neutrinos. We demonstrate that indeed the neutrinos velocity field is locally deflected over time. We translate this signature to the number density of neutrino states that would be detected in a cosmic neutrino detector, revealing the physical nature of the neutrino particle. We update the calculations in~\cite{Long_2014} to compute the capture rate in tritium detectors with the new calculated number density distribution of helicity states, also taking into account the impact of the unkown neutrino mass scale \cite{Roulet:2018fyh} in normal hierarchy \cite{hierarchy} and gravitation clustering.


\section{Method}


\subsection{Cosmic neutrino background. Helicity composition.}

In the hot, dense conditions of the early Universe, neutrinos are in thermal equilibrium with the rest of particles in the plasma. At the time of freeze-out, the neutrino scattering rate drops below the Hubble expansion rate and relativistic neutrinos free stream with a frozen Fermi-Dirac distribution. After this time, neutrino energy and density redshift and the neutrino content decoheres in mass eigenstates. This implies that the content of neutrinos can be predicted with independence of the initial conditions in the Standard model of particle physics and cosmology. At late times, neutrinos (at least two of them) become non relativistic and gravity of large scale structures modifies the distribution of neutrinos.

We will briefly discuss now the helicity content of neutrinos when they become non relativistic. Further details can be found in~\cite{Long_2014}

If neutrinos are Dirac particles, then they have four degrees of freedom per generation, which are commonly labelled as: 
$\nu_L$ left-handed active neutrino; $\bar \nu_R$ right-handed active anti-neutrino; $\nu_R$ right-handed sterile neutrino; $\bar \nu_L$ left-handed sterile anti-neutrino. Dirac neutrinos are distinguished by their lepton number, which is a conserved quantity. The states $\nu_L$ and $\bar \nu_R$ are active in the sense that they interact via the weak interaction. The states $\nu_R$ and $\bar \nu_L$ are sterile in the sense that they interact only via the Higgs boson, i.e., the mass term. Weak interactions modify only the amount of active states in the early Universe, which acquire the abundance $n_{\nu}(z)$, while sterile neutrinos cannot achieve thermal equilibrium with the standard model (SM). For the Dirac case, the expected spin state abundances of the cosmic relic neutrinos are  $n(\nu_L) = n(\bar \nu_R) = n_{\nu} (z)$ and $n(\nu_R) = n(\bar \nu_L) \approx 0$. In a $C_{\nu} B$ detector sensitive to only neutrinos, one of the four states will be counted.

If neutrinos are Majorana particles, lepton number is not conserved, and there is no quantum number that distinguishes neutrino from  anti-neutrino in the non-relativistic limit. The degrees of freedom are commonly labelled as $\nu_L$ left-handed active neutrino; $\nu_R$ right-handed active neutrino; $N_R$ right-handed sterile neutrino; $N_L$ left-handed sterile anti-neutrino. The active neutrinos interact weakly, and both the left- and right-handed states are populated at freeze-out. The sterile neutrinos interact only through the Higgs boson, but now they are expected to be much heavier than even the electroweak scale. As such, they will decay into a Higgs boson and a lepton, and their relic abundance today is zero. For the Majorana case, the expected spin state abundances of the cosmic relic neutrinos are: $n(\nu_L) = n(\nu_R) = n_{\nu} (z)$; $n(N_R) = n(N_L) = 0$. 

After cooling of neutrinos to the non-relativistic regime, helicity and chirality states do not coincide. Then the abundances today would be, for Dirac neutrinos: $n(\nu_{h_L}) = n(\bar \nu_{h_R}) = n_0$; $n(\nu_{h_R}) = n(\bar \nu_{h_L}) \approx 0$, and for Majorana neutrinos: $n(\nu_{h_L}) = n(\nu_{h_R}) = n_0$; $n(N_{h_R}) = n(N_{h_L}) = 0$, where $n_0 = 56$ cm$^{-3}$ and $n(\nu_{h_L})$ is the number density of negative helicity neutrinos, $n(\nu_{h_R})$ is the number density of positive helicity neutrinos, and so on. The total abundance is the same, $6n_0$, in both cases. However, the C$\nu$B contains both left- and right-helical active neutrinos in the Majorana case, but only left-helical active neutrinos in the Dirac case.

If the neutrinos are not exactly free streaming, due to gravitational clustering, but instead they interact, then the helicity can be flipped. This leads to a redistribution of the
abundances in the Dirac case, $n(\nu_{h_L}) < n_0$ and $n(\nu_{h_R}) > 0$ ( $n(\bar\nu_{h_R}) < n_0$ and $n(\bar \nu_{h_L}) > 0$), but no change in the Majorana case since originally $n(\nu_{h_L}) = n(\nu_{h_R})$.

\subsection{Neutrino capture cross-section for a Tritium detector}

$C_\nu B$ neutrinos can be detected by capture processes in a nucleus. In this section we calculate the rate of $C_\nu B$ capture in tritium without averaging over neutrinos spin as to work out the dependence on the incident neutrino helicity, although we assume non-polarization for the electron detection and nuclei. The process is the following
\begin{equation}
\nu_j \,+\, ^3\text{H} \, \longrightarrow \, ^3\text{He} \,+\, e^-
\label{eq:process}
\end{equation}
where the incident neutrino is taken to be in a mass eigenstate $\nu_j$. Due to the low energies involved in the process (\ref{eq:process}), we can safely work in the four-fermions interaction approximation,
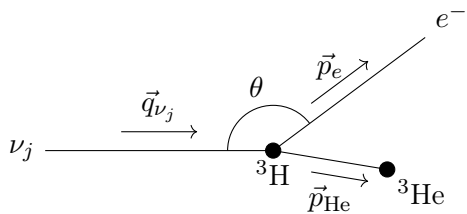
\begin{figure}[H]
\centering
\begin{tikzpicture}
\draw[thin] (-3,0) node[left]{$\nu_j$}--(0,0) ;
\draw[->] (-2,0.25)--(-1,0.25); 
\filldraw(-1.5,0.25) circle(0) node[above]{$\vec{q}_{\nu_j}$};
\draw[thin] (0,0) --(2,1.5) node[above right]{$e^-$};
\draw[thin] (0,0) --(1.5,-0.25) node[below right]{$^3\text{He}$};
\draw[->] (0.5,-0.283)--(1.25,-0.408); 
\draw[->]  (0.5,0.625)--(1.25,1.1875); 
\filldraw(0.75,-0.325) circle(0) node[below]{$\vec{p}_\text{He}$};
\filldraw(0.75,0.8125) circle(0) node[above]{$\vec{p}_e$};
\filldraw(0,0) circle(0.1) node [below]{$^3\text{H}$};
\filldraw(1.5,-0.25) circle(0.1) ;
\draw[thin] (-0.6,0) arc (180:37:0.6);
\filldraw (-0.45,0.85) circle(0) node[right]{$\theta$};
\end{tikzpicture}
\caption{Kinematics in the rest frame of tritium.}
\label{fig:diagram}
\end{figure}
In the rest frame of tritium and taking into account that $p_{\nu_j} = \mathcal{O}(\text{meV})$, the momentum of the outgoing electron is approximately,
\begin{equation}
p_e\simeq\frac{\sqrt{\left(M_H + m_j\right)^4 + \left(m_e^2 - M_\text{He}^2\right)^2 - 2\left(M_H + m_j\right)^2\left(m_e^2 + M_\text{He}^2\right)}}{2\left(M_H + m_j\right)} 
\end{equation}
In this scheme, we calculate the differential cross section not averaged over the incident neutrino spin because depending on which case (Dirac or Majorana) we work, the detector is only sensitive to left-handed neutrinos (Dirac) or both left- and right-handed neutrinos (Majorana).  The differential cross section for a given helicity and mass eigenstates is 
\begin{equation}
\resizebox{0.9\textwidth}{!}{
$\begin{split}
\frac{d\sigma_j (s_\nu,\,q_\nu)}{d\cos\theta}v_{\nu_j}=& \frac{G_F^2 |V_{ud}|^2 |U_{ej}|^2}{4\pi}p_e\, E_e\, F(Z,E_e)\times\\
&\left[\left(\langle f_F\rangle^2 +(g_A/g_V)^2\langle g_{GT}\rangle^2\right)A(s_\nu) + \left(\langle f_F\rangle^2 - \frac{1}{3}(g_A/g_V)^2\langle g_{GT}\rangle^2\right)B(s_\nu) v_e \cos(\theta)\right]
\end{split}
$}
\label{eq:differential cross section}
\end{equation}
where
\begin{subequations}
\begin{equation*}
A(s_\nu) = 1 - 2\,s_\nu v_{\nu_j}
\end{equation*}
\begin{equation*}
B(s_\nu) = v_{\nu_j} - 2\,s_\nu 
\end{equation*}
\end{subequations}

For left-handed neutrinos, $s_\nu = -1/2$ and for right-handed, $s_\nu = 1/2$. In (\ref{eq:differential cross section}), $G_F$ is the weak coupling constant and $V_{ud} = \cos\theta_c$ is an element of the CKM matrix where $\theta_c$ is the Cabibbo angle \cite{ParticleDataGroup:2020ssz}. It is worth nothing that the process is only sensitive to electron neutrinos, that is why we find the PMNS matrix element $U_{ej}$ in (\ref{eq:differential cross section}). $p_e$ ($E_e$) is the outgoing electron momentum (total energy), $F(Z,p_e)$ is the Fermi function where $Z$ is the number of protons of the daughter nucleus, i.e. $^3$He, and $\langle f_F\rangle^2 \approx 0.999$ and $\langle g_{GT}\rangle^2\approx2.720$ \cite{Baroni:2016, De-Leon:2019} are nuclear matrix elements that quantify the probability of finding a neutron in the $^3$H, on which the neutrino can scatter, and a proton in the $^3$He. Finally, the ratio of the axial vector ($g_A$) and the vector ($g_V$) coupling constants is taken to be $1.27641$ \cite{g_ratio} and $\theta$ is the angle between the incident neutrino and the outgoing electron, as shown in Figure \ref{fig:diagram}.

The momentum of the incident neutrinos follows a Fermi-Dirac distribution characterized by the C$\nu$B temperature ($T_\nu = 1.95 $K). Using this and integrating against the scattering angle, one obtains the total capture cross section multiplied by the neutrino velocity, which is the quantity relevant for the capture rate:
\begin{equation}
\sigma_j (s_\nu)\,v_{\nu_j} = \frac{1}{n_0}\frac{4\pi}{(2\pi)^3}\int_0^\infty {dq_\nu\,\frac{q_\nu^2}{1+\exp(q_\nu/T_\nu)}\int_{-1}^{1}{\frac{d\sigma_j (s_\nu,\,q_\nu)}{d\cos\theta}v_{\nu_j}\,d\cos\theta}}
\label{eq:total cross section}
\end{equation}

After integrating over $\theta$ the cross section depends only on $A(s_\nu)$ and no longer on $B(s_\nu)$. The differences in the cross section between the three neutrinos mass eigenstates arise from this term, $A(s_\nu)$. As we will see below, the more non-relativistic the neutrino is, the less velocity it has and, for left-handed, less capture cross section we get. And the other way around for right-handed neutrinos. See second and third columns in Tabla \ref{tab:rates}.

\begin{equation}
\begin{split}
    \sigma_j (s_\nu)\,v_{\nu_j} =& \,\frac{1}{n_0}\frac{4\pi}{(2\pi)^3} \frac{G_F^2 |V_{ud}|^2 |U_{ej}|^2}{2\pi}\times \\
    & \int_0^\infty{ dq_\nu\,\frac{q_\nu^2}{1+\exp(q_\nu/T_\nu)}p_e\, E_e\, F(Z,E_e)\,\left[\langle f_F\rangle^2 +(g_A/g_V)^2\langle g_{GT}\rangle^2\right]A(s_\nu)} \\
    =& \,|U_{ej}|^2\int_0^\infty{ dq_\nu\, \bar{\sigma}(q_\nu) \,A(s_\nu) } = |U_{ej}|^2\int_0^\infty{ dq_\nu\, \bar{\sigma}(q_\nu)\,(1 - 2\,s_\nu v_{\nu_j}) }
\end{split}
\label{eq:A(s)}
\end{equation}
where $\bar{\sigma}(q_\nu)$ is independent of the helicity and also, approximately, of the neutrino mass. 

In (\ref{eq:total cross section}) we have computed the total capture cross section for a given neutrino mass and helicity eigenstates. Calculating the capture rate requires summing over the six possible initial states  ($j = 1,\, 2,\, 3$ and $s_\nu=\pm 1/2$) weighted by the appropriate flux:
\begin{equation}
\Gamma_{C\nu B}=\sum_j\left[\sigma_j(s_\nu=-1/2)v_{\nu_j}n_j(\nu_{hL}) + \sigma_j(s_\nu=+1/2)v_{\nu_j}n_j(\nu_{hR})\right]N_T
\end{equation}
where $N_T$ is the number of targets (100g tritium in this case). If we use the fact that different neutrinos mass eigenstates are equally populated, then
\begin{equation}
\Gamma_{C\nu B}=\left\{ \left[\sum_j\sigma_j(s_\nu=-1/2)v_{\nu_j}\right]n(\nu_{hL}) + \left[\sum_j\sigma_j(s_\nu=+1/2)v_{\nu_j}\right]n(\nu_{hR})\right\}N_T
\label{eq:rate}
\end{equation}
As we have said before, if neutrinos are Dirac particles, in the absence of velocity deflection, we have $n(\nu_{h_L}) = n_0$ and $n(\nu_{h_R}) = 0$, where $n_0 = 56\,\text{cm}^{-3}$; whereas for the Majorana case, $n(\nu_{hL}) = n(\nu_{hR}) = n_0$. The prediction of capture rates is modified in the realistic case of neutrino deflection by the gravitational potential.

To illustrate the dependence on the neutrinos masses, we have calculated the capture rate for three different sets of neutrino masses, all of them in the normal~\cite{hierarchy} ordering ($m_1 < m_2 < m_3$) and constrained by these relations
\begin{subequations}
\begin{equation}
\delta m ^2 = m_2^2 - m_1^2 = 7.5\times10^{-5}\,\text{eV}^2
\end{equation}
\begin{equation}
\Delta m^2 = |m_3^2 - (m_2^2 + m_1^2)/2| = 2.5\times10^{-3}\,\text{eV}^2
\end{equation}
\begin{equation}
m_1 + m_2 + m_3 < 0.20 \, \text{eV}
\end{equation}
\end{subequations}
Table \ref{tab:rates} shows the results of the total cross section multiplied by the neutrino velocity for left- and right-handed neutrinos. It also shows the results of the capture rate on 100g of tritium for the Dirac and Majorana cases. For the three sets of masses, the mass $m_3$ is always non-relativistic while $m_2$ and $m_1$ go from non-relativistic to relativistic. The capture rate for the Dirac case increases as the masses decrease whereas for the Majorana case is the same for all sets of masses because the process does not distinguish between neutrinos or anti-neutrinos, even though the capture is only available for neutrinos. We will adopt as our fiducial case the one in the second group cell from the top. In the first group all neutrinos are non-relativistic. The other cases illustrate the capture rate dependence on the lightest neutrino mass. The table also shows the individual contributions for the different neutrino mass states. The possibility to identify separately the different contributions depends on the detector energy resolution achieved. A more detailed discussion on the experimental resolution achievable and its phenomenological implications can be found on \cite{Ptolemy,Roulet:2020yye}. 

\begin{table}[H]
\makebox[1 \textwidth][c]{       
\resizebox{1 \textwidth}{!}{  
\begin{tabular}{c|c c c c a a}
\hline
\multirow{3}{*}{Neutrinos masses} & \multirow{2}{*}{$\sum_j\sigma_j(\text{left})\cdot v_{\nu_j}$} & \multirow{2}{*}{$\sum_j\sigma_j(\text{right})\cdot v_{\nu_j}$} & \multicolumn{2}{c}{no LSS} & \multicolumn{2}{a}{LSS} \\
& & & $\Gamma_{C\nu B}^D$ & $\Gamma_{C\nu B}^M$ & $\Gamma_{C\nu B}^D$ & $\Gamma_{C\nu B}^M$\\
& $[\times10^{-45}\,\text{cm}^2]$ & $[\times10^{-45}\,\text{cm}^2]$ & $[\text{yr}^{-1}]$ & $[\text{yr}^{-1}]$ & $[\text{yr}^{-1}]$ & $[\text{yr}^{-1}]$\\
\hline
$m_1 = 4.86\cdot10^{-2} \text{eV}$ & 2.5908 & 2.5349 & 2.76 & 5.46 & 2.75 & 5.46 \\
$m_2 = 4.95\cdot10^{-2} \text{eV}$ & 1.1626 & 1.1379 & 1.24 & 2.45 & 1.23 & 2.45 \\
$m_3 = 7.00\cdot10^{-2} \text{eV}$ & 0.0858 & 0.0845 & 0.09 & 0.18 & 0.09 & 0.18\\
\rowcolor{black!15}          TOTAL & 3.8391 & 3.7573 & 4.09 & 8.09 & 4.08 & 8.09 \\
\hline
$m_1 = 1\cdot10^{-3} \text{eV}$ & 3.6799 & 1.4458 & 3.92 & 5.46 & 3.56 & 5.46 \\
$m_2 = 8\cdot10^{-3} \text{eV}$ & 1.2261 & 1.0745 & 1.31 & 2.45 & 1.28 & 2.45 \\
$m_3 = 5\cdot10^{-2} \text{eV}$ & 0.0860 & 0.0842 & 0.09 & 0.18 & 0.09 & 0.18 \\
\rowcolor{black!15}       TOTAL & 4.9920 & 2.6045 & 5.32 & 8.09 & \textbf{4.94} & \textbf{8.09} \\
\hline
$m_1 = 1\cdot10^{-4} \text{eV}$ & 5.0150 & 1.1069 & 5.34 & 5.46 & 4.56 & 5.46 \\
$m_2 = 8\cdot10^{-3} \text{eV}$ & 1.2261 & 1.0745 & 1.31 & 2.45 & 1.28 & 2.45 \\
$m_3 = 5\cdot10^{-2} \text{eV}$ & 0.0860 & 0.0842 & 0.09 & 0.18 & 0.09 & 0.18 \\
\rowcolor{black!15}       TOTAL & 6.3271 & 1.2694 & 6.74 & 8.09 & 5.93 & 8.09 \\
\hline
\end{tabular}
} 
} 
\caption{Summary of cross section and capture rate results. No LSS refers to the standard case when gravitational deflection by cosmological large scale structures is not taken into account and therefore $n(\nu_{h_L})=56 \, \text{cm}^{-3}$. LSS is the case when gravitational deflection by large scale structures is included (computed in this work) and therefore $n(\nu_{h_L})=48 \, \text{cm}^{-3}$. Our reference case is the one shown in the second cell from the top, as suggested by cosmology~\cite{shapefit}.}
\label{tab:rates}
\end{table}

We have anticipated the result of the cosmological Large Scale Structure (LSS) as computed in the following sections to show the final observable capture rates.

Even though $p_e$ ($E_e$) depends on the mass of the neutrino, the dependence is so small that it is not evident in the capture rate. Capture cross section real dependence on the neutrino mass is inside the $A(s_\nu)$ factor, because of the neutrino velocity (\ref{eq:A(s)}). When we sum left and right contributions, neutrino velocity terms cancel and the cross section become effectively independent of the neutrino mass; that is why we get the same total capture rate for the Majorana case for all masses sets, because it is in this case when both left- and right-handed neutrinos have the same density so we sum directly left- and right-cross section and lose the mass dependence (\ref{eq:rate}). 

\subsection{Neutrino Deflection in a Gravitational Potential}

Neutrinos travelling through a gravitational potential will be deflected. This is not the case for their spin. Therefore, gravity has an impact on helicity, the  projected spin onto the direction of movement. 
The clustering of neutrinos within the cold dark matter haloes, i..e, the imprint of large scale structures on neutrinos, has been calculated in multiple semi-analytic and numerical simulations. The neutrino density and peculiar velocity fields have been calculated inside and in the outskirsts of virialized halos, characterizing the neutrino density profiles and the perturbed Fermi-Dirac distribution of neutrino velocities \cite{Villaescusa-Navarro:2012ilf}.

We focus here on an unexploited observable of the neutrino field, the change of direction of the neutrino field since the creation of the halo, which in fact, may lead to a larger signature than previously studied neutrino contributions to the dark matter field. We will calculate numerically this observable and explore the potential to observe it in upcoming and future astronomical surveys. This signature modifies the helicity content of neutrinos and therefore would have an impact on the signal observed in a future $C_{\nu} B$ detector. This is the signature we exploit and that we aim at measuring in the lab by exploiting this signal imprinted in the sky.

As it has been calculated in the last section, the capture rate in a tritium-based detector depends on the neutrino polarized cross section, densities of left- and right-helical neutrinos and the number of targets in the detector (\ref{eq:rate}). For our fiducial case, in the absence of velocity deflection, for the Dirac case, when $n(\nu_{h_L}) = n_0$ and $n(\nu_{h_R}) = 0$, the capture rate expected is 5.32 yr$^{-1}$. Alternatively, for the Majorana case, when $n(\nu_{h_L}) = n(\nu_{h_R}) = n_0$, the capture rate increases to 8.09 yr$^{-1}$. That is an increase of 54\%. The prediction is modified in the realistic case of neutrino deflection by the gravitational potential, as we will demonstrate in the following sections. Gravity changes left-handed neutrinos into right-handed neutrinos; as now $n(\nu_{h_R})\neq0$, the right-handed contribution to the capture rate becomes non-zero in the Dirac case. For non-relativistic neutrinos this effect is very small because left-polarized cross section and right-polarized cross section are almost the same, see second and third columns in Table \ref{tab:rates} for $\nu_3$ neutrinos. Instead, for relativistic neutrinos ($\nu_1$), the change from left-handed to right-handed makes an appreciable effect in the capture rate because the cross section for right-handed relativistic neutrinos is smaller than the cross section for left-handed relativistic neutrinos reducing the total capture rate in the case of velocity deflection by gravity. In our reference case the Dirac capture rate is reduced by gravity from 5.32 to 4.94 yr$^{-1}$. On the other hand, for the Majorana case, we get the same result because neutrino deflection by gravity is the same for left- and right-helical neutrinos so there is not change in the respective densities.


\begin{figure}
\vspace*{1cm}
 \includegraphics[width=\columnwidth]{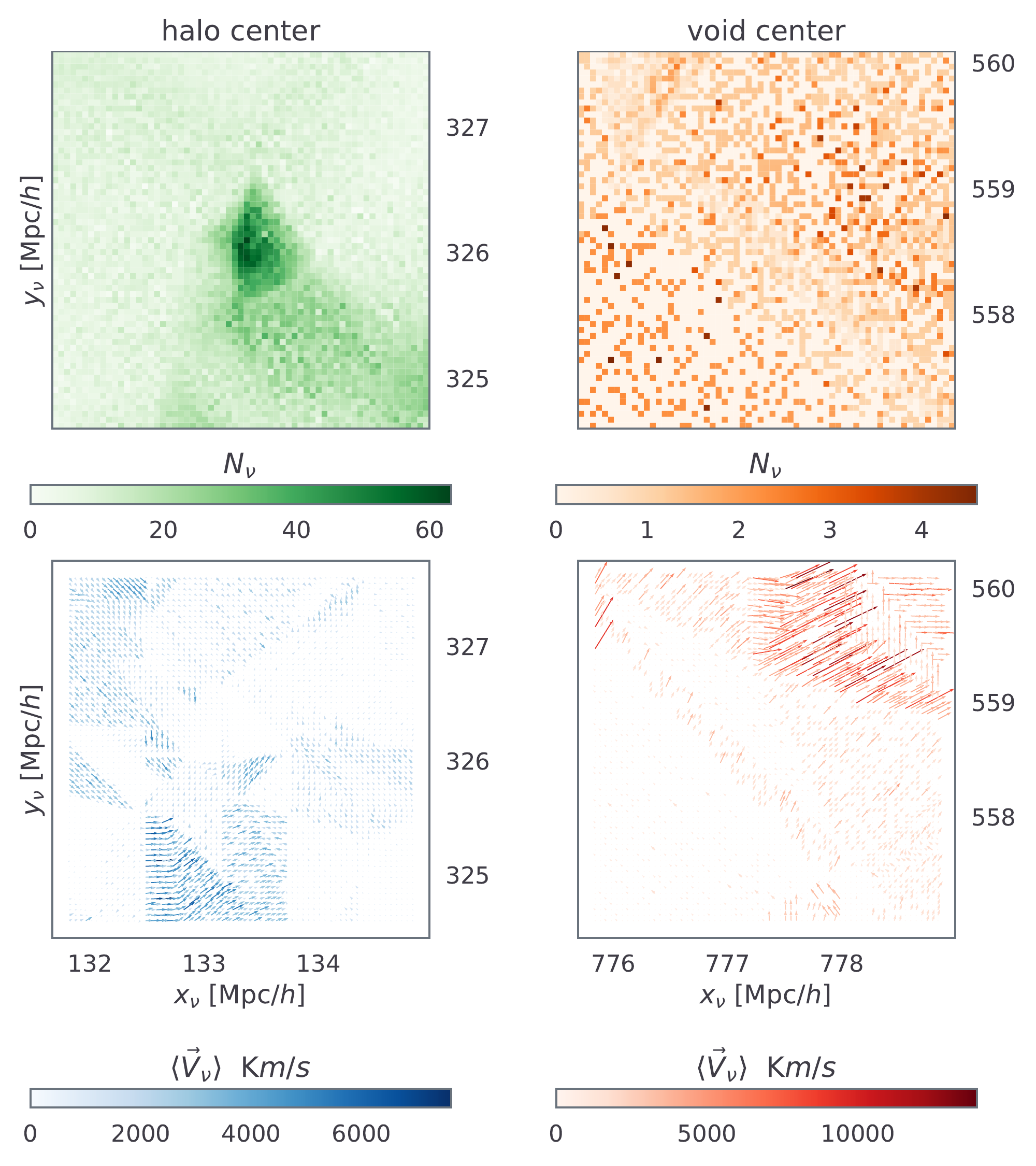}
 \caption{
\label{fig:twoD_section_vel_den}
Central section (thickness $\Delta z = 0.047$ Mpc$/h$) of a 3 Mpc$/h$ side box enclosing the center of one halo (left column) and one void (right column) found in the dark matter plus neutrinos simulation at redshift zero. The top row shows the neutrinos number count in the 64 x 64 pixels, each of volume $V_\mathrm{p}=0.047^3$ $(\mathrm{Mpc}/h)^3$. The bottom row shows the projected 2D average velocities for the same pixels. The color bar reports the 3D velocities magnitude.}
\end{figure}

\begin{figure}
\hspace*{1cm}
 \includegraphics[width=0.8\columnwidth]{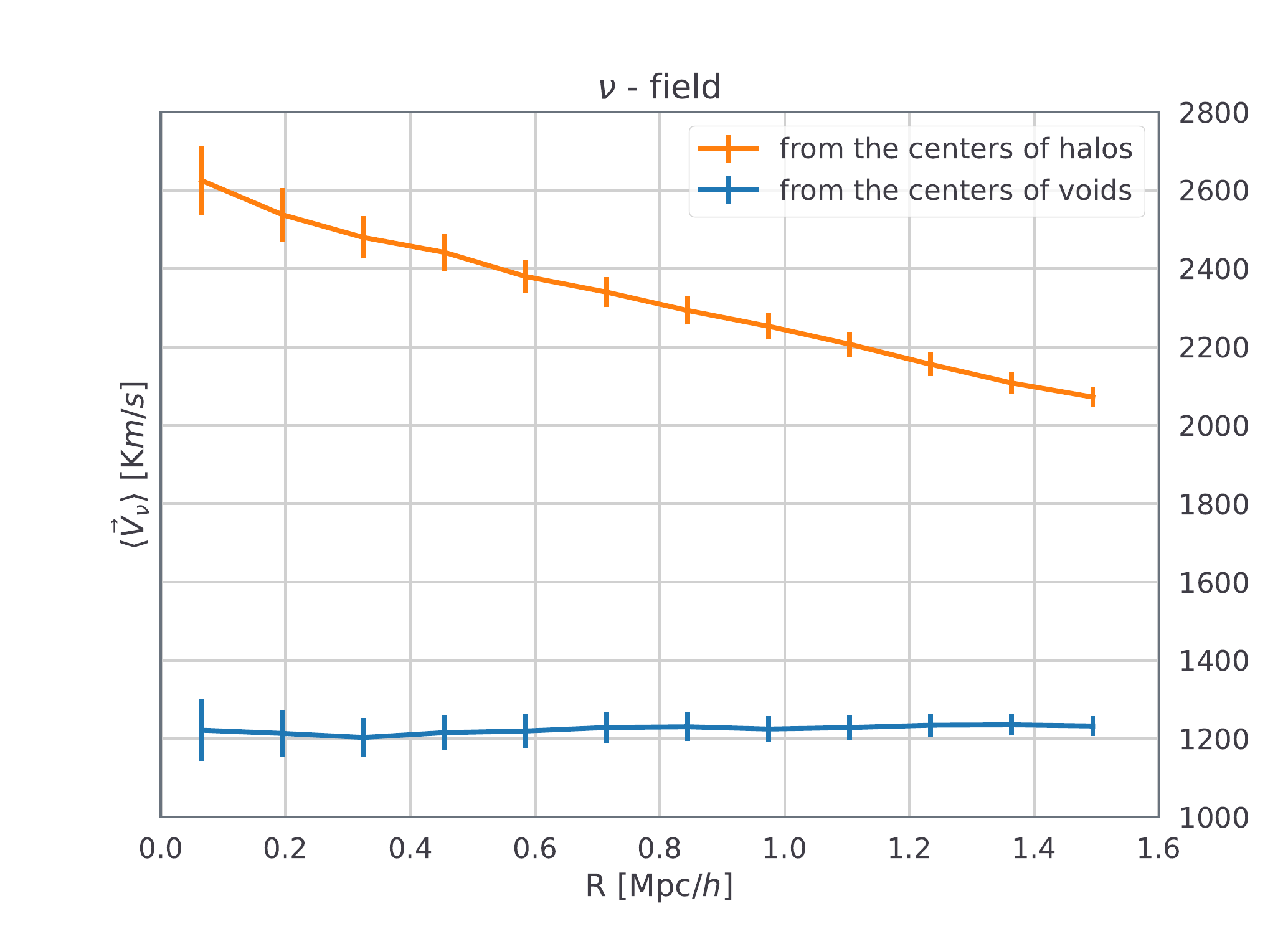}
\caption{
\label{fig:oneD_section_avVel}
Average neutrino velocity field as a function of distance from the halo (void) center obtained from the 100 heaviest halos (largest voids) at redshift zero. The error bars are relative to the mean signal and therefore correspond to the standard distribution divided by the square root of the number of averaged halos (voids). While in voids the average neutrino velocity does not significantly vary, there is an evident decrease in $\langle\Vec{V}_\nu\rangle$ when moving away from the halos centres.}
\end{figure}

\section{Numerical Simulations and Software}
\label{sec:sims_softw}
We want to investigate the change in the velocity field from the moment neutrinos become non-relativistic to $z=0$ in both halos and voids, at least in the ideal case of a numerical simulation. 
To investigate this possibility, we use the N-body simulations from the \textsc{Quijote} suite \cite{Villaescusa-Navarro:2019bje}. The simulations follow the gravitational evolution of $512^3$ dark matter particles, in a periodic cubic box with co-moving size $L=1\, h^{-1}\textrm{Gpc}$. The initial conditions for the simulations were generated at $z=127$ using 2LPT \cite{Springel:2002uv,Crocce:2006ve,Scoccimarro:2011pz}, and we concentrate on one realisation's two different snapshots at redshifts $z=0$ and $z=3$; the latter is the highest redshift output available and we use it as a good approximation to  when neutrinos become non-relativistic as dark matter structure has grown very little between the redhsift at which neutrinos become non-relativistic and $z=3$.   

The underlying cosmology of the \textsc{Quijote} simulations is a flat $\Lambda$CDM model (consistent with the \textit{Planck} satellite CMB analysis  \cite{Planck:2018vyg}). In particular, the matter and baryon density parameters are $\Omega_\mathrm{m}=0.3175$, $\Omega_\mathrm{b}=0.049$, and the dark energy equation of state parameter is $w=-1$; the reduced Hubble parameter is $h=0.6711$, the dark matter fluctuations normalisation parameter is $\sigma_8=0.834$, the scalar spectral index is $n_\mathrm{s}=0.9624$.

Given our purpose to study the deflection  angle of the neutrinos velocity field during a redshift interval, we focus on one of the simulations having massive neutrinos, labelled "Mnu-p" and corresponding to a sum of neutrino masses $M_\nu = 0.1$ eV~\footnote{While the capture rates in Table~\ref{tab:rates} are dominated by less massive neutrinos, this simulation illustrates the effect of deflection for the lower velocity neutrinos, which will be similar for the lower mass case.}. Each simulation of the \textsc{Quijote} suite has both catalogues for the halos and voids, for details on how these are found see \cite{Villaescusa-Navarro:2019bje}. These numerical simulations treat neutrinos as particles thus their velocities are meaningful. 

To measure neutrinos density and velocity fields we use the public DTFE code \cite{Cautun:2011gf} based on Delaunay tessellation for field reconstruction \cite{Schaap:2000se}. This 
method allows switching from discrete samples/measurements to values on a periodic grid maximising the extraction of the information contained in the particle distribution. One of the advantages of this approach is that the derived quantities are volume- and not mass-weighted. This facilitates the comparison with theoretical/analytical quantities which are usually volume-weighted.

\begin{figure}
\hspace*{-1cm}
\includegraphics[width=1.1\columnwidth]{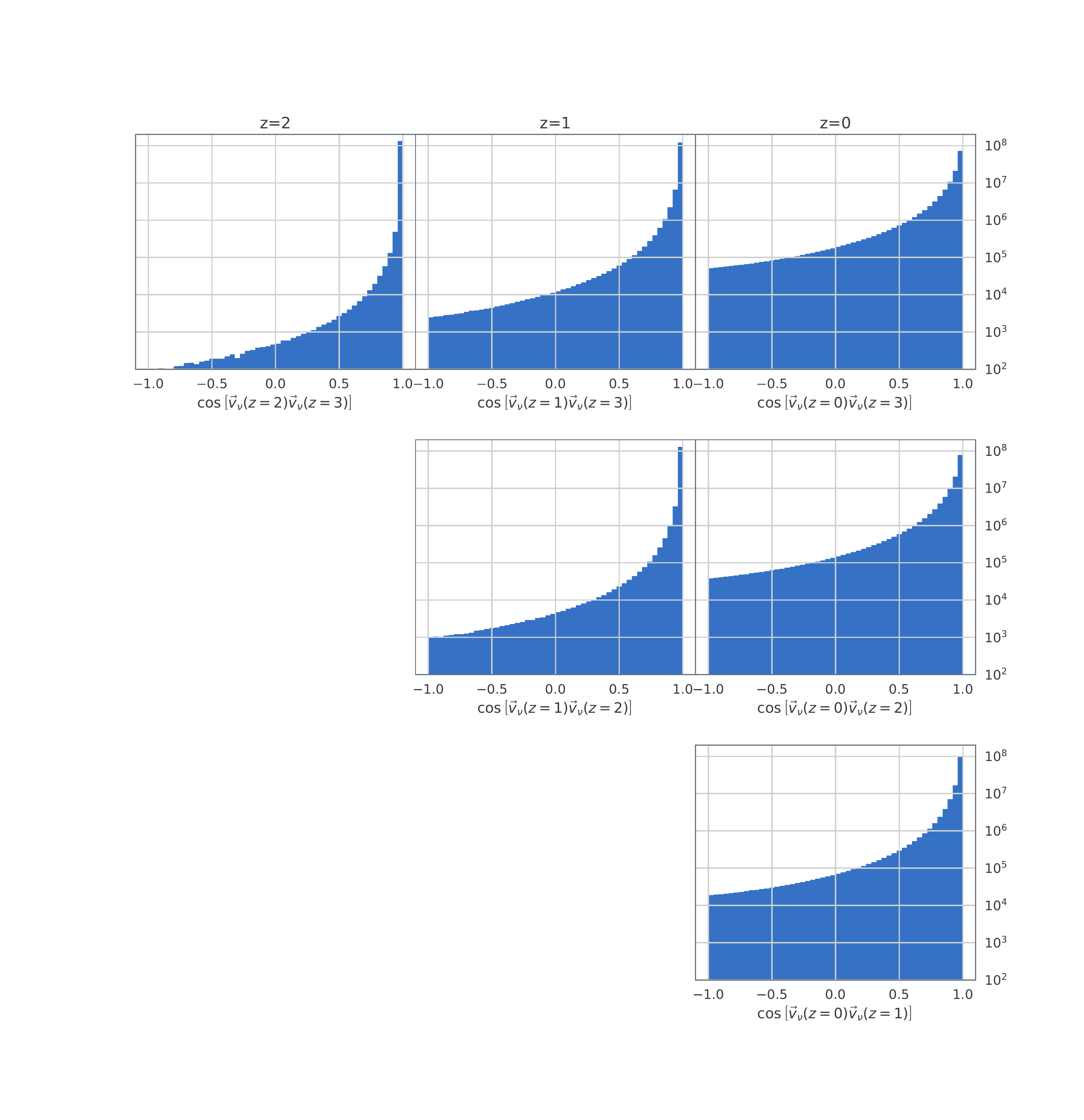}
\caption{
\label{fig:chnagefullbox}
The panels show the number of neutrinos in the whole simulation box as a function of the cosine of the angle between initial and final velocities of each neutrino for different redshift slices. As expected, the number of neutrinos with deflections different from zero grows as non-linear structures develop.
}
\end{figure}

\section{Halos and Voids: a clear signature}

In Figure \ref{fig:twoD_section_vel_den} the derived density and velocity fields using the DTFE code described in Section \ref{sec:sims_softw} are shown in two cases: one halo (left column) and one void (right column). In particular, the cubic box of side 3 Mpc$/h$ is centered at the halo and void respective positions given by the Quijote catalogues. The top row displays the number count of neutrinos in the box's central 2D section made of 64 x 64 pixels of volume $V_\mathrm{p}=0.047^3$ $(\mathrm{Mpc}/h)^3$. In the bottom row we show for the two cases the 2D projected neutrinos average velocities in each pixel, with the respective color-maps describing the 3D vectors' magnitude. 

Figure \ref{fig:twoD_section_vel_den} acts as an illustrative example to visualise the observed quantities in the single case of one halo and one void at redshift zero and to highlight the fact that neutrinos will be deflected by the gravitational field. In order to quantify the gravitational effect on the neutrinos velocity field $\Vec{V}_\nu$, we measure the velocity field around 100 of the most massive halos and inside 100 of the largest voids present in the dark matter simulation. Then we compute the neutrinos average velocity absolute value as a function of the distance from the center of each box. The average signal for both halos and voids is shown in Figure \ref{fig:oneD_section_avVel}. The error bars are given by the standard deviation divided by the square root of the number of considered halos (voids), corresponding to the volume of 100 boxes. The decrease of $\langle\Vec{V}_\nu\rangle$ as the distance from the halo's center becomes larger, is a clear signature of the halo's gravitational field. Indeed $\langle\Vec{V}_\nu\rangle$ remains approximately constant as one moves away from the center of the void. The figure show that the velocity field is stronger closer to the center of a halo.

\begin{figure}
\hspace*{-1.5cm}
\includegraphics[width=1.15\columnwidth]{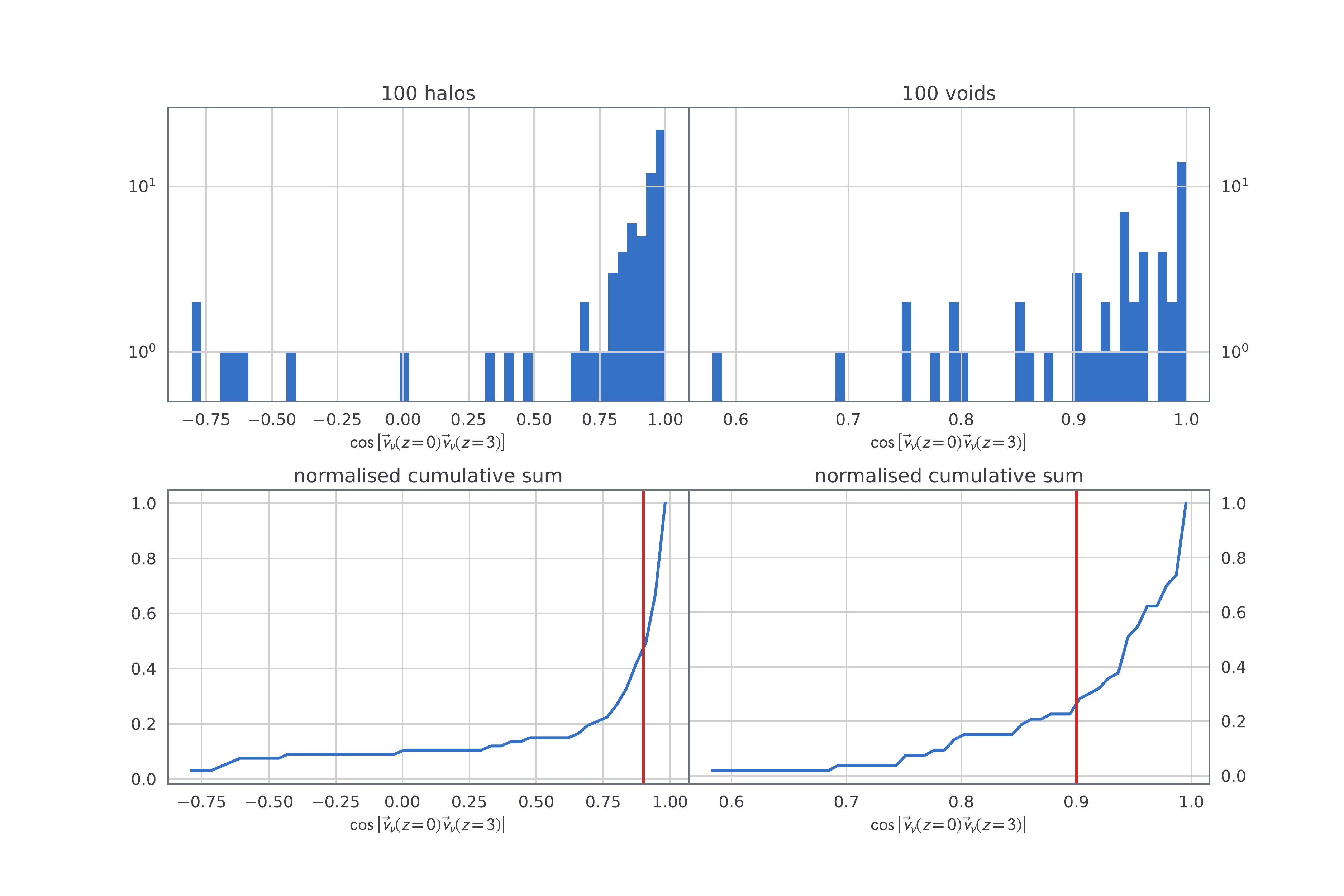}
\caption{
\label{fig:voidshalos}
Top row: neutrino number as a function of deflection angle when averaging the most massive 100 halos and voids in the simulation box. We have integrated in both cases up to $1$ Mpc in radius. Bottom row: normalised cumulative sum. As expected, there is a tendency for the deflection to be less significant in voids than halos.
}
\end{figure}

Given this signature, we can now compute the signal of interest to us, namely: how much the deflection angle for the neutrino changes from the moment the halos/voids just started to collapse, i.e. when the fluctuation is still linear until it is fully virialised. 
To do this, we compute the cosine of the absolute value of the angle between the neutrino velocity at redshift $z=3$ (first output in Quijote simulations) and $z=0$ for each individual neutrino.

Fig.~\ref{fig:chnagefullbox} shows the number of neutrinos for each corresponding angle change of velocity between different redshifts for the whole simulation box. We observe two facts: as expected, since the majority of the Universe is empty, most neutrinos suffer no deflection ($\cos = 1$). The number of deflected neutrinos by gravity grows as non-linear structures grow via gravitational instability. 

We can now focus more closely on the non-linear structures. Using the halo and void catalogues described in the previous section, we produce averages of the deviation angle between $z=3$
 and $z=0$ for the most massive $100$ halos and voids. The distribution of deviation angles is shown in Fig.~\ref{fig:voidshalos}. In this figure we have integrated up to a radius of 1 Mpc.
 As expected, the relative number of neutrinos that get deflected to those that are not has increased compared to the whole box. There is a  difference between halo and voids, with voids producing less deflection as expected.

Fig.~\ref{fig:voidshalos} is our main result. If neutrinos are Dirac we know that the density of neutrinos on Earth, with potential to be captured by weak interations should be $56$ cm$^{-3}$ in the absence of any gravitational effect. The expectation is that this number should be 113 cm$^{-3}$ for Majorana neutrinos. Using numerical simulations we have shown that for Dirac neutrinos there is a change in their density composition on Earth. This factor is in the range $1.05-1.15$ (for voids and halos respectively). So non-linear structures help distinguish the signal between Dirac and Majorana if they are relativistic as they reduce the Dirac signal. 

To illustrate this we also show the redshift evolution of the cosine between neutrino velocity fields in Figure~\ref{fig:chnagefullbox}. Note how as non-linear structure develops, the value of the cosine decreases (i.e. the angle increases).

We can now compute, from Fig.~\ref{fig:voidshalos}, the decrease in the number of left-helical neutrinos expected, if they are Dirac, from their deflection angle, by simply using quantum mechanics probability rule for the change of helicity as
\begin{equation}
    n(\nu_{h_L})_{\rm Dirac} = 56 {\rm cm}^{-3} \left [1 - \sum_{0}^{\pi} [\sin^{2}({\frac{\theta}{2}}) * {\rm weight}(\theta)] / \sum({\rm weight(\theta)}) \right ]
\end{equation}
where weight$(\theta)$ is the number of neutrinos for a given angle. As anticipated above this gives a reduction factor $1.05-1.15$ which corresponds to densities of 54 and 48 cm$^{-3}$ for voids and halos, respectively. As expected, the deflection in a void is less pronounced than in a halo. 

While the Milky Way is in the center of a relatively empty void, it is its massive dark matter halo and that of Andromeda (both are $\sim 10^{12}$ M$_{\odot}$) that will contribute most to the final neutrino velocity deflection. Recall that the Universe is mostly empty and non-linear structures occupy a small volume. Thus a neutrino will most likely be deflected by just one non-linear structure. The signal on Earth should be $n(\nu_{h_L}) = 48$ cm$^{-3}$ and $n(\nu_{h_R}) = 8$ cm$^{-3}$ if neutrinos are Dirac. This has to be compared to $n(\nu_{h_L}) = n(\nu_{h_L})=56$ cm$^{-3}$ if neutrinos are Majorana. A $5\sigma$ detection of the $C_{\nu}B$ at this level should be possible if a $C_{\nu}B$ experiment has a resolution of better than $13$ neutrinos per cm$^{3}$ (or better than one neutrino per year if one considers the capture rate from Table.\ref{tab:rates}) and will provide a proof of the Majorana nature of neutrinos. Note that the effect of gravity is to help widen the difference between the two cases, specially as more relativistic the neutrino is.

\section{Conclusions}

At present time, ground based prototype experiments are being constructed to  measure and characterise the $C_{\nu}B$. This will open a window to when the Universe was just one second old, much earlier than what has been currently explored by the cosmic microwave background.

Here we have proposed, and computed for the first time, a physical effect that should be included in the $C_{\nu}B$ flux estimate of weakly interacting neutrinos. Because the Universe is clustered and non-linear gravitational structures exist in the form of galaxy clusters and voids, neutrinos will suffer a  deflection angle in their passage through the gravitational potential. Gravity produces a change in the helicity content of the $C_{\nu}B$ that modifies the prediction in case neutrinos are Dirac fermions.

We have used numerical simulations from the Quijote suite \cite{Villaescusa-Navarro:2019bje} to show that the gravitational deflection imprinted on the neutrinos velocity field by massive halos/galaxies is detectable. This is reported in  Fig.~\ref{fig:voidshalos}  where the distribution of the average cosine between the neutrino velocity fields at $z=0$ and $z=3$ is shown for halos and voids. This is the signal we propose to be used in ground based $C_{\nu}B$ laboratories.

We have shown that on Earth the effect, using the average signal of $100$ halos, is a decrease on the left-handed neutrinos interacting with the detector of $48$ cm$^{-3}$ instead of $56$ in the  Dirac case: a 15\% decrease. This increases the difference with the case of Majorana neutrinos for which we expect $113$ cm$^{-3}$ ($56$ cm$^{-3}$ for both left- and right-helical neutrinos). We verified the robustness of the result with respect to variations in the LSS environment, halos and voids, and with respect of the redshift at which neutrinos become non-relativistic. 

We have made a precise prediction of what experiments like Ptolemy would measure. The difference of rate observed in for Majorana and Dirac neutrinos depends on the mass scale. Within the current cosmological bound and the evidence of normal hierarchy~\cite{hierarchy}, the rate of Dirac neutrinos capture in a 100 g tritium detector is $4.9^{+1.1}_{-0.8}$ neutrinos per year, which ranges from 0.50 to 0.74 times the rate of Majorana neutrino capture, 8.1 captures per year independently of the helicity composition and mass scale.
Other gravitational affects, like neutrino clustering~\cite{nuhalo,nuhalo2}, have not been discussed here; these, using the scaling formulas in \cite{nuhalo,nuhalo2} will lead to an enhancement of 10\% for both the Dirac and Majorana cases of the neutrino capture rate given the current bounds on the neutrino mass scale~\cite{shapefit}. In an upcoming publication we will show the exact deflection calculation for the solar environment using a  constrained simulation of the Milky Way for light neutrinos ($< 0.001$ eV). In summary, if $C_{\nu}B$ can be detected at $5\sigma$, we conclude that the experiment would distinguish the nature of neutrinos at the $3\sigma$ level.

\begin{acknowledgments}
Funding for the work of RJ was partially provided by project PGC2018-098866- B-I00 MCIN/AEI/10.13039/501100011033 y FEDER ``Una manera de hacer Europa'', and the ``Center of Excellence Maria de Maeztu 2020-2023'' award to the ICCUB
(CEX2019- 000918-M funded by MCIN/AEI/10.13039/501100011033. We thank Davide Gualdi for initial discussions and help with plotting procedures. We also thank Francisco (Paco) Villaescusa-Navarro for advice and correspondence on the QUIJOTE simulations and Esteban Roulet for correspondence on the helicity content and capture rate of the $C_{\nu}B$. 
\end{acknowledgments}


\providecommand{\href}[2]{#2}\begingroup\raggedright\endgroup

\end{document}